\documentclass[reprint,superscriptaddress,amsmath,amssymb,aps,prb]{revtex4-1}

\usepackage{tabu}
\usepackage{graphicx}
\usepackage{epstopdf}
\usepackage{dcolumn}
\usepackage{bm}
\usepackage[breaklinks=true]{hyperref}
\hypersetup{colorlinks=true, citecolor=blue, filecolor=blue, linkcolor=blue,urlcolor=black}
\usepackage{hyperref}
\begin{document}

\preprint{APS/123-QED}

\title{Nonlocal magnon spin transport in yttrium iron garnet with tantalum and platinum spin injection/detection electrodes}

\author{J. Liu}
\thanks{}
\email{jing.liu@rug.nl}
\affiliation{Physics of Nanodevices, Zernike Institute for Advanced Materials, University of Groningen, Nijenborgh 4, 9747 AG Groningen, The Netherlands}
\author{L. J. Cornelissen}
\affiliation{Physics of Nanodevices, Zernike Institute for Advanced Materials, University of Groningen, Nijenborgh 4, 9747 AG Groningen, The Netherlands}%
\author{J. Shan}
\affiliation{Physics of Nanodevices, Zernike Institute for Advanced Materials, University of Groningen, Nijenborgh 4, 9747 AG Groningen, The Netherlands}%
\author{B. J. van Wees}
\affiliation{Physics of Nanodevices, Zernike Institute for Advanced Materials, University of Groningen, Nijenborgh 4, 9747 AG Groningen, The Netherlands}
\author{T. Kuschel}
\affiliation{Physics of Nanodevices, Zernike Institute for Advanced Materials, University of Groningen, Nijenborgh 4, 9747 AG Groningen, The Netherlands}
\affiliation{Center for Spinelectronic Materials and Devices, Department of Physics, Bielefeld University, Universit\"atsstra\ss e 25, 33615 Bielefeld, Germany}%

\date{\today}

\begin{abstract}
We study the magnon spin transport in the magnetic insulator yttrium iron garnet (YIG) in a nonlocal experiment and compare the magnon spin excitation and detection for the heavy metal paramagnetic electrodes  platinum (Pt$|$YIG$|$Pt) and tantalum (Ta$|$YIG$|$Ta). The electrical injection and detection processes rely on the (inverse) spin Hall effect in the heavy metals and the conversion between the electron spin and magnon spin at the heavy metal$|$YIG interface. Pt and Ta possess opposite signs of the spin Hall angle. Furthermore, their heterostructures with YIG have different interface properties, i.e. spin mixing conductances. By varying the distance between injector and detector, the magnon spin transport is studied. Using a circuit model based on the diffusion-relaxation transport theory, a similar magnon relaxation length of $\sim 10\,\mu$m was extracted from both Pt and Ta devices. By changing the injector and detector material from Pt to Ta, the influence of interface properties on the magnon spin transport has been observed. For Ta devices on YIG the spin mixing conductance is reduced compared with Pt devices, which is quantitatively consistent when comparing the dependence of the nonlocal signal on the injector-detector distance with the prediction from the circuit model.
\end{abstract}

\maketitle

Magnons are the quasi-particle representations for collective excitations of spin waves in magnetically ordered systems. Yttrium iron garnet (YIG) is a magnetic insulator with the lowest known magnetic damping at room temperature, which corresponds to the longest magnon lifetime \cite{cherepanov1993saga,barker2016thermal}. Magnons can carry spin angular momentum and in a magnetic insulator, such as YIG, this transport is without moving any electron charge.

Since magnons carry spin information, this makes magnon transport promising for long-range information transmission and processing \cite{chumak2015magnon}. Long-wavelength magnons can be generated by microwave excitation, even as a uniform precession mode with certain GHz-frequency (ferromagnetic resonance, FMR). However, the wavelength of the spin waves limits the size of the smallest devices which can be based on it. In order to develop nano-sized magnon-transport devices, scaling down the wavelength of controllable spin waves becomes attractive for scientists; however, the first step of realizing nano-scaled magnon devices is to use a reliable method to excite and detect short-wavelength magnons in the THz-regime, where exchange interaction dominates. 

One possibility is to inject spins from mobile electrons of a metal which have the energy of $\sim$$\, k_{\textrm{B}}T$ (e.g. $300\,$K$\,\sim$$\,6\,$THz). This has been first theoretically predicted \cite{PhysRevLett.109.096603} by using a ferromagnetic insulator sandwiched by two metallic layers. Here, the spin Hall effect (SHE) \cite{hoffmann2013spin} converts charge current to a pure spin current, while the inverse SHE (ISHE)\cite{saitoh2006conversion} transforms the spin current back to a charge current. Nonlocal magnon spin transport was demonstrated by experiments with Pt$|$YIG$|$Pt lateral structure, where both Pt injector and detector are patterned on top of a YIG film \cite{cornelissen2015long,goennenwein2015non}. Later on, vertical structures Pt$|$YIG$|$Pt(Ta) as proposed in the theoretical paper have been realized \cite{wu2016observation,li2016observation}. 

So far, different spin current sources have been used to create the spin accumulation at the interface of the YIG, either employing the SHE of heavy paramagnetic metals or by the polarized spin current in a ferromagnetic metal associated with the anomalous Hall effect (AHE) \cite{das2017spin}. In both cases, the transferring of the spin angular momentum is based on the scattering between mobile electrons in the metal with the localized electrons in YIG\cite{PhysRevLett.109.096603,PhysRevLett.108.246601,cornelissen2015long}. Therefore, we call it electrical injection of magnon spins (the reciprocal process: detection). Besides the electrical excitation of magnon spins, applying a charge current to a metal bar produces Joule heating. The resulting temperature gradient in the magnetic insulator also generates a magnon spin current. This is called the thermal injection via Joule heating \cite{cornelissen2015long,PhysRevB.94.174437}, also known as nonlocal spin Seebeck effect, in contrast to the electrical injection via SHE. External heaters have also been used to specifically study the thermal generation of the magnon spins, such as laser-heating \cite{giles2015long,giles2017thermally}. Here, the nonlocal transport is governed by both magnons and phonons \cite{giles2015long,giles2017thermally}. Therefore, one should be careful when analyzing the obtained nonlocal spin Seebeck signals, e.g. when determining the magnon spin diffusion length \cite{PhysRevB.96.184427}.

Platinum is so-far the most commonly used spin-Hall metal for building up the nonlocal devices in order to study the magnon spin transport of magnetic insulators such as YIG\cite{cornelissen2015long,goennenwein2015non,giles2015long,PhysRevB.93.020403,wu2016observation,li2016observation,ganzhorn2016magnon,PhysRevB.94.014412,velez2016competing,cornelissen2016temperature,PhysRevB.94.174437,zhou2017lateral,liu2017MPHEMAMR,PhysRevB.96.104441,ganzhorn2017magnon,wesenberg2017long,giles2017thermally,PhysRevB.96.184427,das2017spin}, nickel ferrite (NiFe$_2$O$_4$, NFO)\cite{shan2017nonlocal} and gadolinium iron garnet (Gd$_3$Fe$_5$O$_{12}$, GdIG) \cite{ganzhorn2017non}. The nonlocal magnon spin transport properties have been studied depending on, e.g., magnetic field strength\cite{PhysRevB.93.020403}, temperature\cite{cornelissen2016temperature}, measurement geometries (longitudinal and transverse)\cite{liu2017MPHEMAMR} as well as thickness of the magnetic insulator and transparency of the contacts\cite{PhysRevB.96.104441}. Platinum has a large spin Hall angle ($\theta_{\textrm{SH}}$), a large spin mixing conductance to YIG ($G_{\textrm{Pt}\mid \textrm{YIG}}^{\uparrow\downarrow}$) and short spin flip time ($\tau_{\textrm{sf}}$) \cite{PhysRevB.87.174417}. These properties make Pt a spin-Hall metal with high spin and charge current conversion efficiency as well as a relatively "transparent" contact to YIG in terms of spin current. A material with opposite sign of the spin Hall angle, such as Ta, has not been used so far for lateral experiments. For vertical devices, Ta has been used as one of the electrodes \cite{li2016observation}. 

\begin{figure}[t!]
	\centering
	\includegraphics[width=1\linewidth]{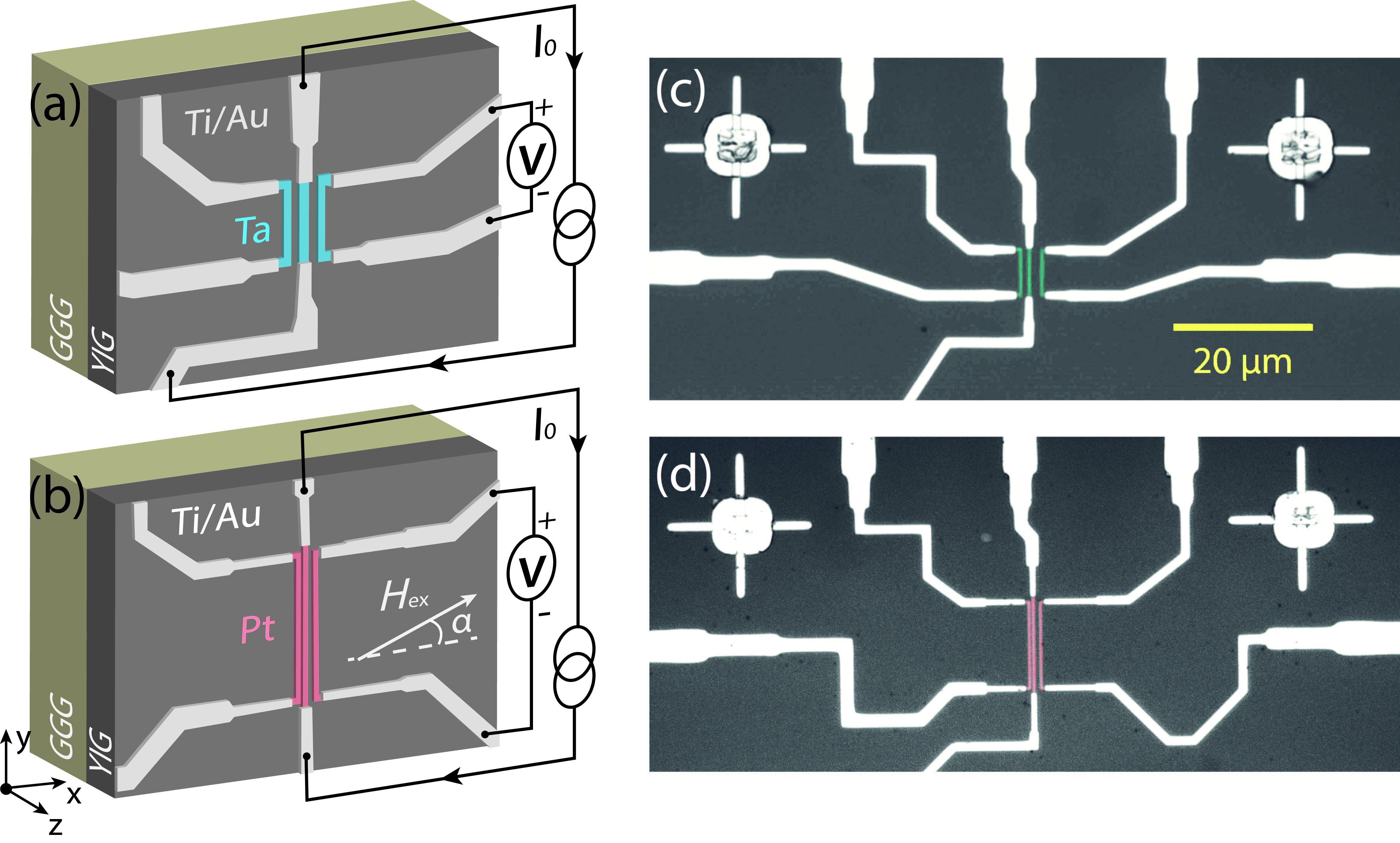}
	\caption{(a,b) Schematic illustration of the typical  device measurement configuration and (c,d) corresponding optical microscope images with false colors for (a,c) Ta devices and (b,d) Pt devices. In an xyz-coordinate system as shown in (a) and (b), YIG thin films (dark grey) lie in the xy-plane on top of GGG substrate (green). Ta (blue) and Pt (pink) bars are along the y-axis. Ti/Au leads (bright grey) are used to connect the device to the electronic measurement setup. We used the measurement configuration shown in (a) and (b): the positive sign of the nonlocal voltage $V_{\textrm{nl}}$ corresponds to a higher voltage potential in the upper lead compared to the lower lead. "+" and "-" symbols around the voltmeter indicate the fashion how the voltmeter is connected. The angle between the external magnetic field $\textit{\textbf{H}}_{\textrm{ex}}$ and the positive x-axis is defined as $\alpha$, which is positive when $\textit{\textbf{H}}_{\textrm{ex}}$ rotates anticlockwisely from $\alpha=0$. }
	\label{OM}
\end{figure}

Here, we present a nonlocal magnon spin transport study with injector and detector both made of Ta on top of YIG. We compare our results with the classical case using Pt. Due to the fact that $\theta_{\textrm{SH}}$ has opposite sign for Pt and Ta\cite{liu2012spin}, we paid special attention to the sign of the SHE-induced nonlocal magnon spin transport and the nonlocal spin Seebeck effect signal. Besides, the influence of the interface properties ($G_{\textrm{Pt}\mid \textrm{YIG}}^{\uparrow\downarrow}>G_{\textrm{Ta}\mid \textrm{YIG}}^{\uparrow\downarrow}$ \cite{PhysRevB.87.174417,vlietstra2014simultaneous}) on magnon spin transport has been investigated.  

The Pt devices with smaller injector-to-detector distance (Series A) are fabricated on 200-nm-thick YIG, which is provided by Universite de Bretagne Occidentale in Brest, France. The Pt devices with longer injector-to-detector distance (Series B) and all Ta devices (Series A1 and B1) are patterned on 210-nm-thick YIG commercially obtained from Matesy GmbH. All the YIG films are grown on gadolinium gallium garnet (GGG) substrate by liquid phase epitaxy method. Both Ta and Pt electrodes are patterned by a three-step electron beam lithography and DC-sputtering. The schematic illustration and optical images of the typical nonlocal devices are shown in Fig.$\,$\ref{OM}. The dimension parameters of all devices are summarized in Table \ref{table_geo}. 

\begin{table}[h!]
	\caption{Geometric parameters of Ta and Pt\cite{cornelissen2015long} electrodes for the distance dependent results in Fig.$\,$\ref{distance}.}
	\begin{ruledtabular}
		\begin{tabular}{cccc}
			 &Length &Width &Thickness \\
			 &($\mu$m)& ($\mu$m)&(nm)\\
			 \hline
			 \\
			 Series A (Pt)&7.5/12.5&0.1-0.15&13\\
			 Series B (Pt)&100&0.3&7\\
		     Series A1 (Ta)&8&0.5/0.6&5\\
			 Series B1 (Ta)&40&1&5\\		
		\end{tabular}
	\end{ruledtabular}
	\label{table_geo}
\end{table}

In the experiment, a low-frequency (usually $\omega/2\pi\sim17\,\textrm{Hz}$) AC-charge current (typically $[I_{0}]_{\textrm{rms}}=80\,\mu$A) is sent through the injector strip as shown in Fig.$\,$\ref{OM}. SHE-generated transverse spin current causes a spin accumulation at the interface between the heavy metal and YIG, which electrically excites magnons in YIG. Simultaneously, the Joule heating of the injector charge current creates a temperature gradient in the YIG, which thermally generates the magnon spin current. The excited magnon spins diffuse towards the detector. By using the lock-in technique, we measured the nonlocal charge voltage at the detector as $V_{\textrm{nl}}$. Then the nonlocal resistances are given by $R_{\textrm{nl}}^{1\omega}=\frac{V_{\textrm{nl}}}{I_{0}}$ and $R_{\textrm{nl}}^{2\omega}=\frac{V_{\textrm{nl}}}{I_{0}^{2}}$ for the first and second harmonic signals, respectively. An in-plane magnetic field $\textbf{\emph{H}}_{\textrm{ex}}$ (typically $\mu_{0}H_{\textrm{ex}}=10\,$ mT) is applied to align the magnetization of the YIG film $\textbf{\textit{M}}$ with an angle $\alpha$ (cf. Fig.$\,$\ref{OM}). We vary $\alpha$ by rotating the sample in-plane under a static magnetic field with a stepper motor. All the measurements are carried out at room temperature.

\begin{figure*}[t!]
	\centering
	\includegraphics[width=0.99\linewidth]{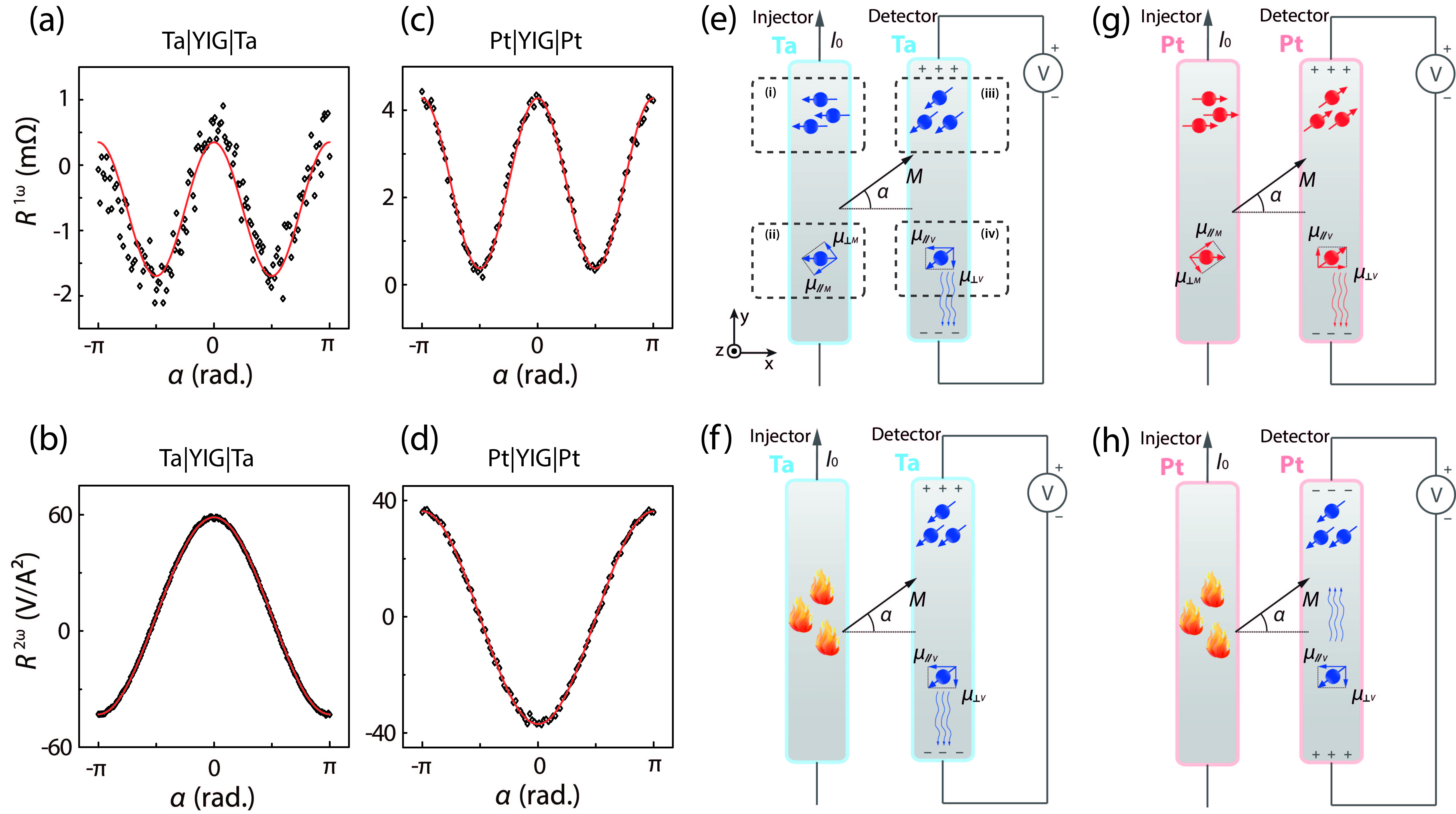}
	\caption{(a)-(d) Magnetic field angle dependent nonlocal measurement results for Ta and Pt devices on YIG thin films. First harmonic signals for (a) Ta and (c) Pt devices with $\cos 2\alpha$-fits (red solid lines). Second harmonic signals for (b) Ta and (d) Pt devices with $\cos \alpha$-fits (red solid lines). The Pt sample is part of series A with injector-detector distance of 540$\,$nm, while the Ta sample is part of Series A1 with injector-detector distance of 850$\,$nm. The offset in the data, e.g. larger nagative offset in (a) and smaller positive one in (b) and (c),  varies from device to device. This is likely caused by the capacitive and inductive coupling between the measurement wires to and from the sample. (e)-(h) Corresponding schematic illustration of the top views of injector and detector at the interface between the heavy metal and YIG. The center-to-center distance between the injector and detector is 850$\,$nm for Ta device and 540$\,$nm for Pt device. In (e), we give a step-by-step procedure of nonlocal magnon injection and detection: (i) SHE of Ta produces spin accumulation  ${\mu}$ with corresponding spin polarization at Ta$|$YIG interface with given $I_{\textrm{0}}$. (ii) Only $\mu_{\parallel\textbf{\textit{M}}}$, which is the component of the spin accumulation along the YIG magnetization $\textbf{\textit{M}}$, can effectively excite magnons. (iii) The excited magnon spin current is detected by the detector strip and converts back to electron spin. (iv) Due to the symmetry of the ISHE, $\mu_{\perp\textbf{\textit{V}}}$, which is the component of the spin accumulation perpendicular to the voltage detection direction, can contribute to the ISHE voltage. With a negative spin Hall angle for Ta, the electron drifts downwards indicated by the arrow of the wave lines. This gives rise to a positive voltage in the given connection configuration of the voltmeter.}
	\label{angle_dependent}
\end{figure*}

Since the electrical magnon injection and detection efficiencies depend on the relative orientation of the electron spin accumulation and the net magnetization of YIG, an angle dependent behavior is expected to be observed as explained in Fig.$\,$\ref{angle_dependent}. Non-equilibrium magnons are generated at the injector, then diffuse towards the detector. For the injection, magnons are excited electrically and thermally. In these two excitation methods, signals scale linearly and quadratically with the excitation current, respectively. For the electrically excited magnons, the SHE-induced electron spin accumulation, i.e. $\mu$, locally introduces the magnons at the heavy metal$|$YIG interface. However, only the eletron spin accumulation parallel to the YIG magnetization, i.e. $\mu_{\parallel \textit{\textbf{M}}}$, can effectively inject magnons. This results in the $\cos\alpha$ angular dependence of the electrical injection via SHE. The non-equilibrium magnon spins with polarization of $\mu_{\parallel \textit{\textbf{M}}}$ diffuse away from the injector under a gradient of the magnon chemical potential\cite{PhysRevB.94.014412}. At the detector, they are converted back to the electron spin current and measured as a voltage signal under an open circuit condition via the ISHE. However, due to the symmetry of the ISHE only the electron spin accumulation with polarization perpendicular to the voltage detection direction, i.e. $\mu_{\perp \textit{\textbf{V}}}$, contributes to the measured voltage. This gives rise to $\cos\alpha$ angular dependence of the electrical detection via ISHE. Therefore, we expect a $\cos2\alpha$ dependence for the total electrically injected nonlocal magnon spin transport (cf. Fig.$\,$\ref{angle_dependent}(e,g)). In the case of thermal excitation, a Joule heating caused temperature gradient is used to generate the magnons. This process is independent from the magnetization direction $\alpha$. The thermally excited magnons are converted back to an electron spin accumulation at the detector again by the ISHE (cf. Fig.$\,$\ref{angle_dependent}(f,h)). Thus, a $\cos\alpha$ dependence is expected for the total thermally injected nonlocal magnon spin transport. The signals produced by electrically and thermally excited magnons can be differentiated by employing the first and second harmonic measurement of a lock-in system, respectively. 

\begin{figure*}[t!]
	\centering
	\includegraphics[width=0.99\linewidth]{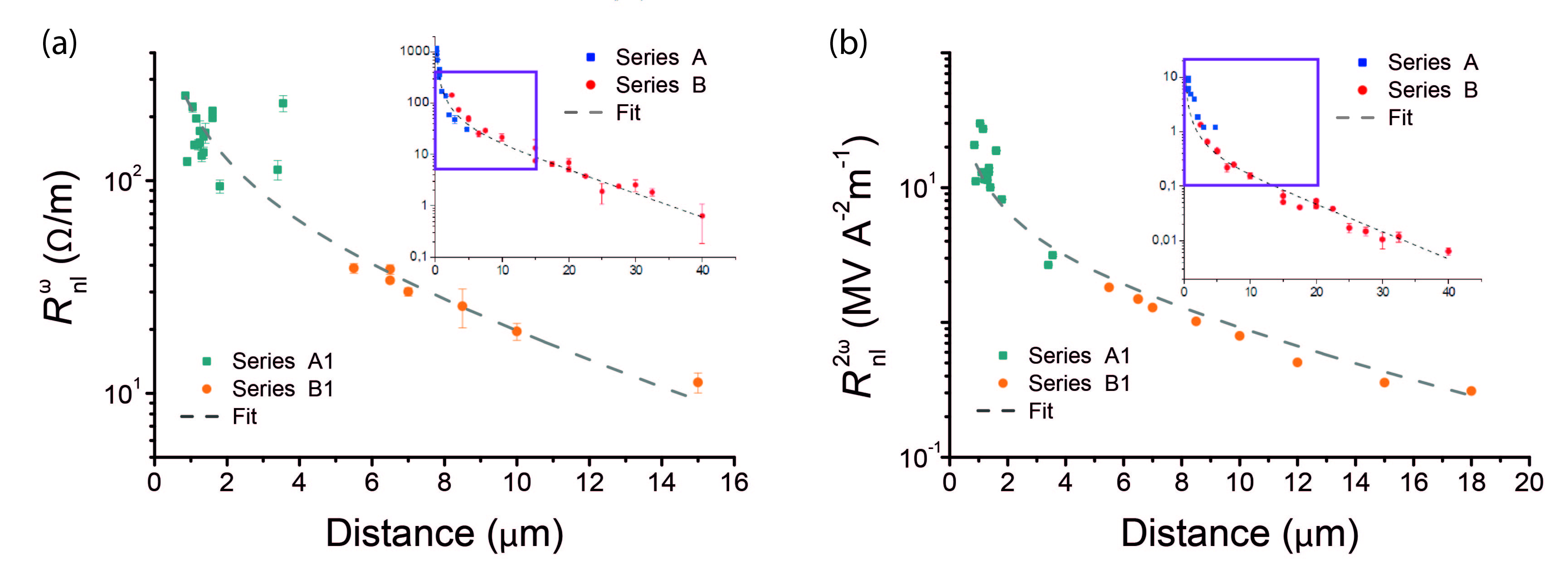}
	\caption{The magnitude of nonlocal resistance of Ta devices as a function of injector-to-detector spacing: (a) first and (b) second harmonic signals  are scaled by the length of the device on logarithmic scale. To compare, in the corresponding inset Pt device data are shown from Cornelissen et al. \cite{cornelissen2015long}. The purple square indicates the same regime as for the Ta data. Each data point is an in-plane magnetic field angle rotation measurement of one device with certain spacing between injector and detector. The magnitude of the first and second harmonic signals are extracted, respectively. Dark cyan and orange squares represent two series of devices fabricated in two times on two pieces of YIG films. The error bar represents the standard error from the $\cos\alpha$ and $\cos2\alpha$ fitting of the first and second harmonic measurement. Note that for the largest injector-to-detector spacing device (L = 18 $\mu$m), the magnitude of the first harmonic signal is smaller than the noise. Therefore, we could not extract the magnitude of the first harmonic signal. However, the second harmonic signal is still measurable. The error bars in (b) are smaller than the size of the data points. Grey lines represent the 1D diffusion-relaxation model fitting. Magnon relaxation length has been extracted with magnitude of $\lambda^{1\omega}_{\textrm{Pt}} =9.4 \pm 0.6\,\mu$m and $\lambda^{2\omega}_{\textrm{Pt}}= 8.7 \pm 0.8\,\mu$m for Pt \cite{cornelissen2015long}. As for Ta, from the data of Series B1 based on Eq.$\,$(\ref{eq_Rc}) we extracted $\lambda^{1\omega}_{\textrm{Ta}} =9.5\pm1.3\,\mu$m and $\lambda^{2\omega}_{\textrm{Ta}}= 9.6\pm1.8\,\mu$m. Due to the large spread of data in Series A1, we fixed these $\lambda_{\textrm{Ta}}$ values to fit the first and second harmonic signals in the whole $L$-range as indicated by the grey dashed lines.}
	\label{distance}
\end{figure*}

The results are presented in Fig.$\,$\ref{angle_dependent}(a)-(d). For both Ta and Pt devices we obtain a $\cos2\alpha$ dependence in case of electrical injection (first harmonic signal, cf. Fig.$\,$\ref{angle_dependent}(a,c)). The sign of the angular dependencies is the same, although Ta and Pt have opposite spin Hall angle. Since there are two SHE processes during injection and detection of the magnon spins, any negative sign of the spin Hall angle for both injector and detector is canceled out and will not affect the sign of the overall electrically excited nonlocal signal. However, the sign of the $\cos\alpha$ dependence in case of thermal injection (second harmonic signal, cf. Fig.$\,$\ref{angle_dependent}(b,d)) changes from Ta to Pt. Since the thermal injection is based on Joule heating and not on the SHE, only one SHE process during the detection is part of the thermally excited nonlocal magnon spin transport. Thus, the sign of the spin Hall angle of the detector material governs the sign of the overall angular dependence of the thermal signal in Fig.$\,$\ref{angle_dependent}(b,d) and the second harmonic signal of the Ta devices is consistently of opposite sign compared to the second harmonic signal of the Pt devices. 

We need to mention that we compare the Ta and Pt devices with comparable injector-to-detector distances when we discuss the sign of the signals. In our previous study with Pt$|$YIG structures, we find that the second harmonic signal, such as the one shown in Fig.$\,$\ref{angle_dependent}(d), changes sign when the injector and detector being very close to each other due to the contribution of the bulk spin Seebeck effect\cite{PhysRevB.94.174437}. The injector-detector distance at which the sign of the second harmonic signals reverses can be different for Ta and Pt devices due to different interface properties between Ta$|$YIG and Pt$|$YIG. However, for the distance dependence of the Ta device we investigated devices with injector-to-detector distances up to 18$\,\mu$m and observed only one sign for the second harmonic signals. Therefore, the sign reversal of the Ta devices, which can be found at smaller injector-to-detector distances, is below the distances probed in this study and is beyond the scope of this paper. The results in Fig.$\,$\ref{angle_dependent} from Pt and Ta devices both have injector-to-detector distance above the sign reversal distance.

\begin{figure}[h!]
	\centering
	\includegraphics[width=1\linewidth]{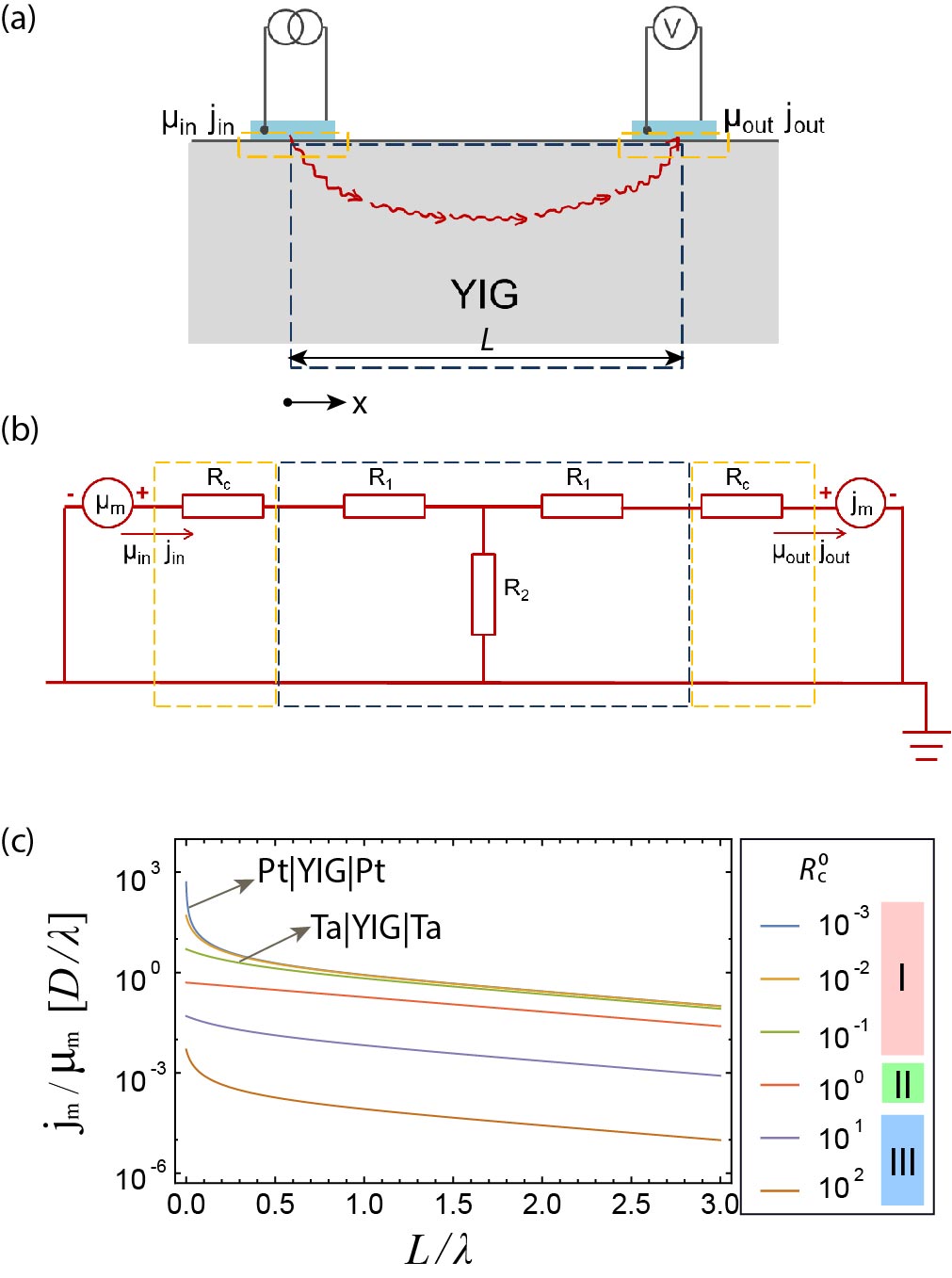}
	\caption{Equivalent circuit model for a one-dimensional diffusive magnon spin transport depending on the contact resistance. (a) Schematic illustration of the cross-section for a typical device. We assume that the magnon spin transport is along x-axis from injector to detector in the YIG channel, which is indicated by the square of dashed line in dark blue. The square of the orange dashed lines represent the contact resistance between the heavy metals and YIG. The distance between the injector and detector is $L$. $\mu_{\textrm{in}}$, $j_{\textrm{in}}$, $\mu_{\textrm{out}}$ and $j_{\textrm{in}}$ are the input and output of the magnon spin electrochemical potentials and magnon spin current densities. (b) The corresponding equivalent circuit. $\mu_{\textrm{m}}$ suggests that the magnon chemical potential is built up at the injector, resulting in a magnon chemical potential gradient in the 1D-YIG channel, like a "magnon voltage source". This gives rise to the magnon spin current. $j_{\textrm{m}}$ represents that at the detector the magnon spin current with densities of $j_{\textrm{m}}$ is measured, working as a "magnon ammeter". $R_{1}$ and $R_{2}$ are the resistances associated with the magnon spin diffusion and relaxation process, respectively. (c) The magnitude of output magnon spin current density $j_{\textrm{m}}$ normalized by input electrochemical potential of magnon spin accumulation $\mu_{\textrm{m}}$ as a function of $L/\lambda$ with different relative contact resistances $R_{c}^{0}$. }
	\label{model}
\end{figure}

Distance-dependent behaviors of the first and second harmonic signals for the Pt devices have recently been obtained by Cornelissen et al. \cite{cornelissen2015long} as shown in Fig.$\,$\ref{distance} (insets). From both first and second harmonic signals, two regimes have been observed: diffusion- and relaxation-dominant magnon spin transport in Series A and B, respectively. The 1D diffusion-relaxation model has been employed to discuss the magnon spin transport behavior. The nonlocal resistance $R_{\textrm{nl}}$ is proportional to the magnitude of the output magnon spin current density
$j_{\textrm{m}}=-D\frac{\partial{\mu_{\textrm{m}}}}{\partial{x}}$ ($\frac{\partial{\mu_{\textrm{m}}}}{\partial{x}}$ is the gradient of the magnon spin chemical potential c.f. Fig.$\,$\ref{model}) and described by
\begin{equation}
R_{\textrm{nl}}\sim j_{\textrm{m}}(x=L),
\end{equation}
\begin{equation}
R_{\textrm{nl}}\sim \frac{C}{\lambda}\frac{e^{\frac{L}{\lambda}}}{1-e^{\frac{2L}{\lambda}}},
\label{eq:nonlocalfitting}
\end{equation}
 
where $x=L$ indicates $j_{\textrm{m}}$ at the detector, C is a fitting parameter and $\lambda$ is the magnon diffusion length ($\lambda=\sqrt{D\tau}$, $D\,$: the magnon diffusion constant, $\tau\,$: the magnon relaxation time). The measured nonlocal voltage $V_{\textrm{nl}}$ results from the magnon spin current $j_{\textrm{m}}$ being converted via the ISHE at the detector. The nonlocal resistance, $R_{\textrm{nl}}$, is obtained by normalizing $V_{\textrm{nl}}$ with the current we send through the injector strip. By fitting the first and second harmonic data with Eq.$\,$(\ref{eq:nonlocalfitting}), similar magnon relaxation lengths have been extracted for the first and second harmonic signals with magnitude of $\lambda^{1\omega}_{\textrm{Pt}}=9.4\pm0.6\,\mu$m and $\lambda^{2\omega}_{\textrm{Pt}}=8.7\pm0.8\,\mu$m. This indicates that the same magnons are excited via electrical and thermal injection, although the injection mechanism (either electrical or thermal) is different.

For Ta devices, both injection and detection electrodes are tantalum. In contrast to the distance-dependent behavior of Pt devices shown in the inset of Fig.$\,$\ref{distance}, the first harmonic signals of the Ta devices do not drop obviously in the short L regime (Series A1) as shown in Fig.$\,$\ref{distance}(a). The $\frac{1}{L}$ character of the diffusive decay is suppressed. In the large L regime (Series B1), they already start to decay exponentially. Moreover, in terms of the magnitude of the signals, the length-scaled first harmonic signals for Ta devices are comparable with that of Pt devices. However, the magnitude of the length-scaled second harmonic signals for Ta are around 5 times larger than that of Pt. This also points to the influence of the electrode resistance (the resistivity of Ta is almost one order of magnitude larger than that of Pt). Therefore, we expect the Pt$\mid$YIG interface to have a larger spin mixing conductance compared with the Ta$\mid$YIG interface according to the following circuit model. 

The one-dimensional diffusion-relaxation model that has been applied to explain the magnon spin transport in YIG for the Pt devices\cite{cornelissen2015long} is now used to describe the magnon spin transport in YIG for the Ta devices. Therefore, we changed the description of magnon spin injection/detection electrodes from Pt to Ta and introduced a higher spin resistance at the interface of electrode/YIG in the equivalent resistor circuit model, which is described by a contact resistance $R_{\textrm{c}}$ (cf. Fig.$\,$\ref{model}). In order to understand the different distance-dependent behavior of Ta and Pt devices, we vary the contact resistance $R_{\textrm{c}}$ in the model as shown in Fig.$\,$\ref{model}. The input and output magnon spin current density ($j_{\textrm{in}}$ and $j_{\textrm{out}}$) and the electrochemical potential ($\mu_{\textrm{in}}$ and $\mu_{\textrm{out}}$) are sketched in Fig.$\,$\ref{model}(a) and represented by the magnon spin transport parameters ($\lambda$, $L$ and $D$):
\begin{equation} 
\begin{pmatrix}
j_{\textrm{in}} \\
j_{\textrm{out}}
\end{pmatrix}
\\=\frac{D}{\lambda(e^{\frac{L}{\lambda}}-e^{-\frac{L}{\lambda}})}
\begin{pmatrix}
K & -2 \\
2 & -K)
\end{pmatrix}
\begin{pmatrix}
\mu_{\textrm{in}} \\
\mu_{\textrm{out}}
\end{pmatrix},\label{eqn_physics}
\end{equation}
where $K=e^{\frac{L}{\lambda}}+e^{-\frac{L}{\lambda}}$. After obtaining the relation between the "voltage" and "current" of the 1D magnon spin transport channel, we build up an equivalent circuit as shown in Fig.$\,$\ref{model}(b). It consists of two diffusive resistors $R_{1}$ in series and one relaxation-related resistor $R_{2}$ being parallel with one of the diffusive resistors depending on the direction of the magnon spin current. At the injector, a magnon accumulation of $\mu_{\textrm{m}}$ is built up, which is the starting point of the magnon spin transport channel. At the detector, the magnon spin current is measured as $j_{\textrm{m}}$. Here, Kirchhoff's circuit laws were applied to obtain the relation between the "voltage" and the "current" of the magnon spin transport in terms of the equivalent resistors (first assume $R_{\textrm{c}}=0$):
\begin{equation}
\begin{pmatrix}
j_{\textrm{in}} \\
j_{\textrm{out}}
\end{pmatrix}=\frac{1}{R_{1}^{2}+2R_{1}R_{2}}
\begin{pmatrix}
Q & -R_{2} \\
R_{2} & -Q
\end{pmatrix}
\begin{pmatrix}
\mu_{\textrm{in}} \\
\mu_{\textrm{out}}
\end{pmatrix},\label{eqn_circuit}
\end{equation}
where $Q=R_{1}+R_{2}$.
Equalizing Eqs. (\ref{eqn_physics}) and (\ref{eqn_circuit}), equivalent resistors $R_{1}$ and $R_{2}$ can be expressed by the magnon spin transport parameters as

\begin{equation}
R_{1}=\frac{\lambda}{D}\frac{e^{\frac{L}{\lambda}}-e^{-\frac{L}{\lambda}}}{e^{\frac{L}{\lambda}}+e^{-\frac{L}{\lambda}}+2},
\label{eq_R1}
\end{equation}

\begin{equation}
R_{2}=\frac{2\lambda}{D}\frac{e^{\frac{L}{\lambda}}-e^{-\frac{L}{\lambda}}}{(e^{\frac{L}{\lambda}}+e^{-\frac{L}{\lambda}}+2)(e^{\frac{L}{\lambda}}+e^{-\frac{L}{\lambda}}-2)},
\label{eq_R2}
\end{equation}
from which we simply take $\lambda/D$ as the unit of the equivalent resistance. Now if we consider contact resistance being non-zero ( $R_{\textrm{c}}\neq0$ ), we obtain 

\begin{equation}
\frac{j_{\textrm{m}}}{\mu_{\textrm{m}}}=\frac{R_{2}}{(R_{1}+R_{\textrm{c}})(R_{1}+R_{\textrm{c}}+2R_{2})},
\label{eq_Rc}
\end{equation}
where we can write the contact resistance as
\begin{equation}
R_{\textrm{c}}=\frac{\lambda}{D}R_{\textrm{c}}^{0}
\end{equation}
with the same unit of $\frac{\lambda}{D}$ as for $R_{1}$ and $R_{2}$. We call $R_{\textrm{c}}^{0}$ relative contact resistance. In Fig.$\,$\ref{model}(c), we show the distance dependent behavior of the nonlocal magnon signals predicted by this equivalent circuit model with different magnitude of $R_{\textrm{c}}^{0}$. According to the feature of the distance-dependent behavior, we can classify it into three regimes of magnon spin transport in terms of the magnitude of $R_{\textrm{c}}^{0}$. Firstly, for the more "transparent" contact in regime I (e.g. $R_{\textrm{c}}^{0}=10^{-3}$), a $\frac{1}{L}$-decay is observed in the diffusive regime ($\frac{L}{\lambda}\lesssim1$). In the relaxation regime ($\frac{L}{\lambda}\gtrsim1$), the output signal decays exponentially. This is also what is expected from the fully transparent contact model\cite{cornelissen2015long}. In this regime (I), increasing $R_{\textrm{c}}^{0}$ will decrease the $\frac{1}{L}$-decay character in the diffusive regime, while does not have much effect on the magnon spin transport in the relaxation regime. In regime II with less "transparent" contact, the diffusive decay behavior is significantly suppressed (e.g. $R_{\textrm{c}}^{0}=10^{0}$). Moreover, the overall strength of the signal, $\frac{j_{m}}{\mu_{m}}$, will be suppressed. However, the slope of the exponential decay stays the same, which corresponds to the magnon relaxation length. In regime III with really large $R_{\textrm{c}}$, the $\frac{1}{L}$-decay behavior in the diffusive regime is introduced again. It only appears in this model with confined magnon spin transport channel in between the injector and detector, which is irrelevant to the magnon spin transport with our device geometries.

By comparing the Pt and Ta results with the model, the behavior of the Ta devices is consistent with the equivalent circuit modeling for more "transparent" contacts (cf. Fig.$\,$\ref{model}(c)), still being less "transparent" than the Pt contacts. Thus, increasing $R_{\textrm{c}}$ by changing the materials from Pt to Ta decreases the $\frac{1}{L}$-decay character while it does not suppress the overall magnitude of the output as shown in the first harmonic signals. 

The fitting curve for the Ta devices shown in Fig.$\,$\ref{distance}(a) is fitted properly to the data. The extracted magnon relaxation length is $\lambda^{1\omega}_{\textrm{Ta}}=9.5\pm1.3\,\mu$m. However, there is a large spread of the data points in Series A1. This might be related to the fact that the first harmonic signals are more sensitive to the interface resistance, since the contact resistance is involved in both injection and detection. More data points with shorter injector-to-detector separation (less than 1 $\mu$m) should be detected in the future to further confirm the magnon spin transport behavior in the diffusive regime for Ta devices. 

For the second harmonic signals (cf. Fig.$\,$\ref{distance}(b)) a magnon spin diffusion lengths of $\lambda^{2\omega}_{\textrm{Ta}}=9.6\pm1.8\,\mu$m is extracted by fitting the data with the same formula. Compared with the first harmonic signals, the $\frac{1}{L}$-diffusive decay character is more notable. This might be owing to the different origin of the first and second harmonic signals. For the first harmonic signals the electrically excited spin current has to pass through the Ta$\mid$YIG interface twice at injection and detection. Nevertheless, in terms of the second harmonic signals, the thermally excited magnons do not need to go across the interface at the injector. Therefore, this might be the reason that the second harmonic signals are less sensitive to the interface resistance.   

To conclude, we compare the distance-dependent nonlocal magnon spin transport of Ta$|$YIG$|$Ta and Pt$|$YIG$|$Pt devices. In both case, two regimes of the diffusion and relaxation dominant magnon spin transport are observed. Similar relaxation lengths ($\sim$ 9-10\,$\mu$m) are extracted for first and second harmonic signals from Ta devices for the electrically and thermally excited magnons, respectively. These results are comparable with their counterpart Pt devices\cite{cornelissen2015long}. The spin Hall angle of Ta and Pt is of opposite sign. However, the angular dependence of the electrical injection has the same sign for both Ta and Pt devices, because of having two spin Hall processes involved. The angular dependence of the thermal injection changes sign from Ta to Pt devices. Here, only one spin Hall process is involved during the detection of the thermally excited magnon spin transport. By changing the material from Pt to Ta, the spin mixing conductance at the interface reduces. This suppresses the diffusive decay character in the diffusive regime while it does not affect the overall magnon spin output signals. By comparing the magnon spin transport behaviors of Ta and Pt to the equivalent circuit model, we found that the contact resistance of Ta$|$YIG$|$Ta is less "transparent" with respect to Pt$|$YIG$|$Pt. Moreover, compared to the second harmonic signals, first harmonic signals are shown to be more sensitive to the contact resistance introduced by changing the probing material from Pt to Ta. This is attributed to the different magnon excitation origins. So far we study our experimental observation based on the diffusion model; however, new physics models, such as hydrodynamic viscosity theory, can be developed to understand and predict the magnon spin transport in the insulating system in the future as for the electron transport in the conducting system \cite{scaffidi2017hydrodynamic}.
 
We acknowledge H. M. de Roosz, J. G. Holstein, H. Adema and T. J. Schouten for their technical assistance. This work is part of the research program Magnon Spintronics (MSP) No. 159 financed by the Netherlands Organisation for Scientific Research (NWO). We also appreciate NanoLab NL and the Zernike Institute for Advanced Materials. This research is partly financed by the NWO Spinoza prize awarded to Prof. B.J. van Wees. Further support by EU FP7 ICT Grant No. 612759 InSpin and the Deutsche Forschungsgemeinschaft (DFG) within the priority program Spin Caloric Transport (SPP 1538, KU3271/1-1) is gratefully acknowledged.

\bibliography{reference}

\end{document}